# A Paradigm Shift in Catheter Development: Thermally Drawn Polymeric Fibers for MR-Guided Cardiovascular Interventions

Mohamed E. M. K. Abdelaziz[1,2]†, Libaihe Tian[1,3]†, Thomas Lottner[4], Simon Reiss[4], Timo Heidt[5], Alexander Maier[5], Klaus Düring[6], Constantin von zur Mühlen[5], Michael Bock[4], Eric Yeatman[2], Guang-Zhong Yang[7], Burak Temelkuran[1,3,8]*

[1]The Hamlyn Centre for Robotic Surgery, Imperial College London; London, SW7 2AZ, UK.

[2]Department of Electrical and Electronic Engineering, Faculty of Engineering, Imperial College London; London, SW7 2AZ, UK.

[3]Department of Metabolism, Digestion and Reproduction, Faculty of Medicine, Imperial College London; London, SW7 2AZ, UK.

[4]Department of Diagnostic and Interventional Radiology, Medical Physics, Faculty of Medicine, University of Freiburg; Freiburg, 79106, Germany.

[5]Department of Cardiology and Angiology, University Heart Center Freiburg - Bad Krozingen, Faculty of Medicine, University of Freiburg; Freiburg, 79106, Germany.

[6] MaRVis Interventional GmbH; Garmisch-Partenkirchen, 82467, Germany.

[7]Institute of Medical Robots, Shanghai Jiao Tong University; Shanghai, 200240, China.

[8]The Rosalind Franklin Institute; Didcot, OX11 0QS, UK.

*Corresponding author. Email: b.temelkuran@imperial.ac.uk

†These authors contributed equally to this work.

**Abstract:** Cardiovascular diseases (CVDs) and congenital heart diseases (CHD) pose significant global health challenges. Fluoroscopy-guided endovascular interventions, though effective, are accompanied by ionizing radiation concerns, especially in pediatric cases. Magnetic resonance imaging (MRI) emerges as a radiation-free alternative, offering superior soft tissue visualization and functional insights. However, the lack of compatible instruments remains a hurdle. We present two novel catheter systems—a tendon-driven steerable catheter and an active tracking Tiger-shaped catheter—fabricated using a unique fiber drawing technique. These catheters, showcasing mechanical properties similar to commercial counterparts, have undergone rigorous in-vitro and in-vivo testing, yielding promising outcomes. This innovative approach has the potential to streamline medical device development, thus enhancing patient care in MR-guided interventions.

**One-Sentence Summary:** Thermally drawn catheter showcased in MR-guided cardiovascular intervention on a porcine model.

**Main Text**

## INTRODUCTION

Cardiovascular diseases (CVDs) account for approximately 17.9 million deaths annually, standing as the primary global cause of mortality *(1)*. Concurrently, congenital heart disease (CHD) is particularly notable, ranking among the most prevalent congenital disorders and impacting 0.8% to 1.2% of live births worldwide *(2, 3)*. In the treatment of CVD, fluoroscopy-guided endovascular interventions have emerged as a prominent choice due to their minimally invasive nature and effectiveness in reducing hospital stays. However, the management of complex CHD often requires multiple palliative or corrective procedures throughout a patient's lifetime, necessitating repeated fluoroscopic catheterizations for both diagnostic and interventional purposes.

Despite its established status as the "gold standard" for endovascular interventions *(4)*, fluoroscopy, which utilizes X-ray imaging with iodinated contrast agents, poses significant safety concerns. The ionizing radiation involved carries substantial risks for both patients and medical professionals, with the radiation exposure being associated with a notable cancer risk (1:137 to 1:370) *(5)*. These concerns are particularly acute in pediatric cases *(6)*. Additionally, fluoroscopically guided cardiac catheterization, hindered by poor soft-tissue contrast, presents challenges in the precious positioning of interventional tools. Operators frequently rely on previously acquired contrast angiographic images or mental reconstruction of anatomical structures due to these constraints. Such limited visualization can prolong procedures, increase X-ray exposure, and elevate the risk of complications. Furthermore, the use of radiopaque contrast agents for indirect vessel visualization can potentially compromise kidney function *(7)*.

Magnetic resonance imaging (MRI) has attracted attention as a radiation-free alternative to address these issues *(8)*. MRI offers distinct advantages, including its unparalleled capacity to visualize soft tissues *(9, 10)* and provide functional information such as blood flow, tissue oxygenation, diffusion, perfusion, and mechanical properties (elastography) *(11, 12)*. While gadolinium-based contrast agents, commonly used in MR-guided endovascular interventions, have an excellent safety profile compared to iodine-containing contrast agents for X-ray imaging, they may cause severe side-effects in patients with impaired renal function *(13, 14)*. However, recent studies have demonstrated MRI's capability for performing angiography and endovascular interventions with minimal or no contrast agent *(14)*. The recent development of novel imaging sequences, with up to 20 images per second *(15, 16)*, has paved the way for real-time MR-guided interventions. Early feasibility studies have explored applications such as right heart catheterization *(17)*, electrophysiology ablations *(18, 19)*, endomyocardial biopsy *(20)* and coronary interventions *(21)*.

Despite its promising potential, MR-guided interventions are hindered by the lack of a broad range of instruments that are compatible with MRI environment, including guidewires, sheaths, and catheters *(11, 22)*. Most commercially available instruments are tailored for X-ray-guided procedures and rely on metallic materials, which often lead to image artifacts that can obscure essential anatomical information in the vicinity of the instruments. Elongated metallic structures such as guidewires can also cause potentially dangerous radiofrequency-induced heating under an MRI environment *(23)*, especially, when their length approaches the resonance length of the radio-frequency fields in tissue *(24)*. Polymers are considered MR-safe alternatives to metals, however, directly replacing them comes with challenges in magnetic visibility, mechanical performance, and manufacturing. Despite significant effort devoted to the development of compatible, i.e. MR Safe or MR Conditional according to ASTM F2503-23 *(25)*, instrumentation and tracking and visualization of instruments, there are still various challenges that limit the commercialization of these devices and the popularity of MR-guided interventions *(26)*.

The development of fabrication approaches that enable fast and affordable prototyping is essential for accelerating the research around these devices. This paper outlines a novel, highly-customized prototyping method for a rapid and cost-effective catheter fabrication that employs fiber drawing technology—a well-established and scalable technique. The thermal fiber drawing process has emerged as a pivotal tool for creating high-aspect ratio devices with multi-material and multifunction across multiple applications. The process commences with a macroscopic preform of single or multiple materials, typically several centimeters in diameter and tens of centimeters in length, shaped to replicate the desired cross-sectional geometry of the final fiber. Through thermal softening and drawing in a viscous state, the preform attains the desired diameter and length while preserving transverse geometry throughout the fiber.

Although high aspect ratio devices are prevalent in the medical field, the application of thermal drawing for fabricating surgical instruments has only been explored to a limited extent. Abdelaziz et al. pioneered the use of thermally drawn catheters for endovascular interventions *(27, 28)*. Leber et al. integrated several functions into a miniaturized thermally drawn robotic fiber and demonstrated the feasibility of such robotic fibers in a phantom model for endovascular interventions *(29)*. Additional applications include a phantom study on auripuncture and drug delivery using a fiber-based steerable robot *(30)*, as well as the creation of a programmable bevel-tip needle for neurosurgery through thermal drawing technology *(31)*. These developments not only showcase the significant advantages of thermal drawing in miniaturizing complex structures but also highlight its great potential in medical device manufacturing.

This paper focuses on addressing the lack of instruments in MR-guided endovascular interventions using the innovative fiber drawing technique. It culminates in the development of two novel catheter systems: a tendon-driven steerable catheter and a Tiger-shaped catheter with active tracking. The integration of steerable features aims to address the difficulty of selecting from a multitude of passive catheter shapes, sizes, and features, as well as their limited maneuverability. Both catheter systems, designed specifically for navigation, were successfully demonstrated *in vitro* with a phantom and *in vivo* using a porcine model. It is worth noting that the entire fabrication process was completed within a few weeks at our dedicated research facility. This cost-effective and highly customized fabrication approach holds great potential, empowering researchers and companies with increased flexibility to explore a wide range of designs and modifications without incurring extensive investments or prolonged production timelines.

## RESULTS

### Fabrication of multi-lumen tubing using thermal drawing technique

To address the lack of compatible instruments for MRI-guided cardiovascular interventions, we developed two catheter systems: a tendon-driven steerable catheter and a Tiger-shaped catheter *(32)* with active tracking (Fig. 1, D to E). These catheters are built around the core component of multi-lumen tubing, which are fabricated using our proposed thermal drawing technique. In contrast to commonly used extrusion, which has limitations such as extrudate swelling and complex mold design *(33)*, thermal drawing proved to be highly effective and cost-efficient for producing high aspect ratio devices. In addition, we introduced two features to further enhance the functionality of the catheters: a preform twist for the steerable catheter and wire feeding for the Tiger-shaped catheter, which will be elaborated on in the following sections.

To carry out the thermal drawing process, a custom-draw tower equipped with a three-zone tube furnace was employed (Fig. 1C). The preforms were carefully fed into the furnace and subjected to controlled temperature profiles, including preheating, heating above the glass transition temperature, and quenching (Fig. 1B). Real-time monitoring of the drawn fibers' outer diameter and tension using a laser micrometer and three-wheel tension sensor enabled precise adjustments to the draw parameters, ensuring the production of high-quality fibers.

For steerable catheters, polycarbonate (PC) and polyetherimide (PEI) were identified as suitable candidates, due to their excellent mechanical property and biocompatibility. The fabrication process commenced with the design and production of macroscopic preforms tailored to accommodate the desired components. In case of the steerable catheters, the preforms were crafted with five precisely positioned (fig. S1A) through drilling holes (Fig. 1A). The resulting catheter (Fig. 1D) was then drawn from the preform, with the diameter scaling from 20 mm to approximately 2 mm. During the fabrication process, the cross-sectional geometry of the preform was retained, while the diameter of the drilling holes was reduced about 10-fold from 6 mm to around 0.6 mm, forming uniformly elongated channels.

In the case of Tiger-shaped catheter, cyclic olefin copolymer elastomer (COCe) was selected as the material for the tip and PC for the shaft. The design of this catheter shaft consists of a cross-sectional configuration featuring a narrow lumen for wire and another wider lumen allowing for the smooth passage of other instruments. The intended functionality was achieved by drilling holes through the off-the-shelf polymer rods as preforms, thermally drawing them into fibers with specified diameter. After drawing, the resulting catheter shaft (Fig. 1E) possessed a diameter of 2.5 mm and featured two circular channels—one with a diameter of 0.2 mm, securely holding the wire, and another with a diameter of 1.9 mm to accommodate the passage of additional instruments.

Our integral approach offers a notable advantage in the rapid and cost-effective manufacturing of high aspect ratio medical devices. By repurposing the thermal fiber drawing process, the need for complex tooling or manufacturing processes was eliminated, allowing for greater innovation in design and material selection. This approach not only accelerates research and development but also promotes the use of minimally invasive procedure. Notably, each catheter can be drawn from scratch within a matter of hours, demonstrating the efficiency and feasibility of our method. Furthermore, our thermal drawing approach offers additional benefits beyond rapid tubing fabrication, as it seamlessly integrates the creation of helical lumina structure and wire feeding in a single step, a feat difficult to achieve with traditional approaches.

**Multi-lumen tubing with helical lumina structure**

Undesired motion in flexible pull-wire driven instruments, which are characterized by three phenomena (explained in detail in fig. S4), that have often been overlooked in the literature due to manufacturing limitations. To address these issues, Bogusky et al. *(34)* proposed a solution that incorporated helically routed side lumina (pull-wires) along the catheter shaft to distribute compressive loads and balance bending moments. They presented a multi-step fabrication method involving main and supplemental mandrels fed through a rotating nose cone in a helical pattern.

Inspired by Bogusky et al.'s work, we propose a new approach for fabricating multi-lumen tubing with helically routed channels in a single step, to suppress the undesired motion caused by pull-wire tension in our steerable catheter. This method utilizes the thermal drawing process, wherein the preform is spun while drawing to create axially asymmetric fibers. Previous studies have explored spinning preforms to fabricate various types of fibers, including twisted single-mode fibers *(35)*, helically twisted photonic crystal fibers *(36)*, and fiber-based probes for brain activity mapping *(37)*. As shown in the Fig. 2A, a spinning motor is attached to the linear stage, enabling the preform to spin at an adjustable angular velocity during the draw. This approach allows the fabrication of long lengths of multi-lumen tubing with helical channels, offering pitch sizes ranging from millimeters to tens of centimeters. Additionally, the versatility of this approach extends to creating both straight and helical lumens by modifying the spinning motor, as demonstrated in Fig.2B with resultant multi-lumen tubing featuring helical and straight lumens.

To assess the performance of our helically routed lumina, we conducted finite element analysis (FEA) and examined the friction between the pull-wires and the helically routed channels. Preliminary investigations indicated that helical pitches of 50 mm and 60 mm minimized friction for one-meter-long tubing. Using ANSYS 2020 R2, we created simplified catheter models with fixed pull-wires to analyze the uncontrolled displacement of the catheter tip (fig. S5). Results showed a significant reduction in maximum uncontrolled displacement at the flexible distal tip, decreasing by 80% from 1.05 mm to 0.20 mm when helically routed pull-wires were employed. Bench-top studies were also conducted to evaluate the performance of the helically routed pull-wires in compensating for passive deflections of the catheter shaft.

**Steerable catheter system**

By implementing the aforementioned thermal drawing approaches, the drawn multi-lumen tubing of different configurations (table S1) was processed and assembled into steerable catheter. The system comprised two main components: a catheter body with a steerable tip and a manual handle for remote tip control (Fig. 3A). We implemented a tendon-driven steering mechanism with an MR-visible tendon *(38)* serving as the pull-wire, providing unobstructed MR-visibility from the handle to the distal end. To ensure complete safety, the entire catheter system was constructed using MR-safe materials *(25)*.

The catheter body featured a flexible yet stiff shaft (around 90 cm) and a more flexible distal end (around 10 cm). It included a central working channel for other instruments and liquid injection, and four peripheral channels for pull wires, enhancing both functionality and visibility. To reduce fabrication time and cost, this design allowed for scalable manufacturing, maintaining consistent cross-sectional geometry throughout the length. Braid reinforcement was strategically incorporated along the length to enhance mechanical properties such as torquability, whip reduction, and kink resistance without significantly increasing the catheter's dimensions. As for the steerable tip (Fig. 3B), inspired by Clogenson et al.'s monolithic approach *(39, 40)*, we employed laser cutting to enhance the flexibility and maneuverability of the distal end, with 2 DOF perpendicular compliant flexure hinge joints *(41)* formed through notches in the tubing. The catheter body is then covered with two layers of polyethylene heat shrink tubing to encapsulate the braid reinforcement and prevent thrombogenesis. An atraumatic tip made of low shore hardness platinum-cured silicone tubing is added, followed by a hydrophilic polymer coating to reduce friction and improve maneuverability. To enhance visibility in MR images, we incorporate passive negative markers with iron microparticles *(21)* at the distal end and proximal tip of the catheter.

The steering of the catheter's distal end was simplified through an intuitive control method, eliminating the need for complex hardware or software *(42)*. A carefully designed handle works as the interface, achieving remote steering of the catheter tip and performing other functions (Fig. 3C). Similar to the ACUSON AcuNav™ catheter, the 3D-printed handle comprised two dials positioned on both sides, each deflecting the distal end in two directions (1 DOF) for four-way bending. This is achieved through a double-rack and pinion mechanism that converts rotary motion into linear motion. As the dial rotates, the pinion and racks move in opposite directions, tensioning one pull-wire while releasing the other. This results in the flexible distal end bending towards the tensioned pull-wire. The involute contours of the dial provide an improved grip, and the gear-like profile allows for seamless integration with other gears, enhancing the handle's potential for direct integration with robotics. A pin vice is mounted at the handle's proximal end for fine-tuning the tensioning of the pull-wires during assembly. Additionally, a universal Luer lock port is positioned at the rear of the handle to facilitate fluid injection and the insertion of a microguidewire into the catheter's central lumen.

## Multi-lumen tubing with wire feeding

For the Tiger-shaped active tracking catheter, we exploited another unique capability of thermal drawing. Through the fusion of the thermal drawing process with a feeding mechanism, wires can be integrated into the fiber *(29, 43)*, a task often considered challenging with similar mass production techniques. As shown in Fig. 2C, the wire is threaded through designed channels within the cross-sectional structure and co-fed with the preform. As a result, the channels constrict along with the entire cross-sectional configuration, securely anchoring the fed wire within the fiber. For a detailed visualization of the co-feeding procedure, please refer to the supplementary video S7 in *(44)*.

## Active tracking catheter

The Tiger-shaped catheter, composed of PC material, yields poor MR imaging visualization, necessitating the addition of an active tracking system to enhance visibility especially at the catheter tip. This active tracking system consists of a receive radiofrequency (RF) coil resonant at the Larmor frequency of the MRI system (128 MHz at $B_0$ = 3 Tesla). Here, a single-loop coil (18*1.6 mm²) etched from 35 µm copper on a 50 µm polyimide (Kapton) was added to the tip of the catheter (Fig. 8C). This coil has been successfully used in previous studies of MR-guided coronary catheterization *(21, 45, 46)*. The coil requires an electrical connection via a coaxial cable to the MRI receiver system *(47, 48)*. To guide the cable along the interior of the catheter, during the fabrication process of the Tiger-shaped catheter, a coaxial wire with a diameter of 0.2 mm is co-fed into the smaller channel of the preform, forming a fiber with an embedded wire.

## Bench-top mechanical characteristic for steerable catheters

To thoroughly evaluate the mechanical performance of our self-fabricated catheter shafts in comparison to commercially available counterparts, we conducted a comprehensive set of experiments. The evaluation focused on key characteristics, namely flexibility (flexural rigidity), pushability (axial rigidity), and torquability (torsional rigidity). The results of flexural rigidity (Experiment 1-1), axial rigidity (Experiment 2) and torsional rigidity (Experiment 3-1) were integrated and compared in Fig. 4A. Catheter shaft samples, with specific details provided in table S1, S2 and S3, were subjected to testing using custom-designed setups. Additionally, we evaluated the performance of the assembled catheter handle system in terms of steerability and examined the suppression effect of the helically routed pull-wires on undesired motion of the flexible tip during passive bending.

Flexibility reflects the stiffness of the catheter when subjected to bending and is crucial for guiding and supporting intravascular devices while maintaining distal flexibility. Figures S6, A and C displays the detailed results from the conducted experiments. Experiment 1-1 involved bending the catheter shafts and measuring the force required. Under dry conditions, our in-house fabricated catheter shafts exhibited higher flexural rigidity than commercial catheters. PEI non-braided shafts with parallel and helical lumina showed flexural rigidity ranging from 877.2 Nmm$^2$ to 974.6 Nmm$^2$, while PC shafts with parallel lumina ranged from 532.8 Nmm$^2$ to 653.1 Nmm$^2$. After immersion in water (Experiment 1-2), the flexural rigidity of our catheters decreased but remained comparable to commercial catheters. Stress relaxation over time (Experiment 1-3) was observed, with the Medtronic™ Guiding Catheters experiencing the largest drop (averaging 22.5%), while our PC catheter shafts with helical lumina showed the smallest drop (averaging 8.10%).

Pushability, which measures the catheter's ability to transmit forces in the axial direction, was assessed in Experiment 2, with results presented in fig. S7, A and D. Catheters with higher flexural rigidity demonstrated superior axial force transmission. The PEI catheter shafts exhibited the highest axial rigidity, ranging from 149.07 N to 177.39 N. Over-braiding increased the axial rigidity by 37.06% for parallel lumina and 10.17% for helical lumina in PC catheter shafts. With the exception of the braided PC catheter shaft with parallel lumina, our in-house fabricated catheters displayed pushability comparable to commercial catheters.

The torquability of the catheter was investigated in Experiments 3-1 and 3-2. In Experiment 3-1 involved twisting the samples around their axis using a servo motor, and catheter shafts with parallel lumina exhibited higher torsional rigidity compared to helical lumina shafts. The detailed results of the experiments can be seen in fig. S8A. Braiding improved torsional rigidity by approximately 71.6% for parallel lumina and 65.9% for helical lumina. Some commercial catheters experienced kinking due to thin walls (fig. S8C). In Experiment 3-2, the torque response of the distal end was examined in passively bent catheters simulating the transfemoral trajectory of catheter across the aortic arch. As illustrated in Fig. 4B, catheter shafts with higher flexural rigidity showed poor linear torque transmission, resulting in catheter whip motion. PC-based catheters and certain commercial catheters demonstrated smoother and linear torque transmission. Overall, our in-house fabricated catheter shafts exhibited comparable torsional rigidity and torque response to commercial catheters.

The steerability of integrated catheter-handle assemblies was evaluated in Experiment 4-1, during which the workspace and positioning accuracy of the catheter were assessed, and Experiment 4-2 evaluated the steering performance after passive bending. The results of Experiment 4-1 and 4-2 are graphically illustrated in fig. S13 to fig. S28. Catheters demonstrated good repeatability, with standard deviations of less than 1 mm for most samples, which is similar to what has been reported in previous work *(49)* and less than 5 mm overall. Backlash, indicating the difference in positioning between forward and backward steering, was observed in all samples. Catheters with straight lumina had an average backlash of about 5 mm, while the catheter with helical lumina had a larger backlash of about 10 mm. Backlash increased when straight lumina catheters were bent, but this was not observed in the helical lumina catheter. The torques required to turn the knob fell within the range of ±0.02 Nm, indicating the forces involved in catheter deflection and assembly friction.

Experiment 5 tested the effectiveness of helically routed pull-wires in suppressing passive deflections. In Fig. 4C, the helical lumina catheters demonstrated much smaller deflections compared to parallel lumina catheters, ranging from 0.3 mm to 0.4 mm versus 5.6 mm to 12.4 mm, respectively. These differences in deflection between the parallel lumina samples magnitude can be attributed to diameter variations. For instance, PEI A, with a larger outer diameter (2.7 mm to 2.88 mm), displayed greater deflections compared to PC A and PEI B, which had outer diameters ranging from 2.41 mm to 2.55 mm. It is worth noting that larger tip deflections can negatively impact catheter steering accuracy and potentially cause harm to surrounding tissues and vessel walls.

### *In vitro* MR Compatibility and Phantom Study for Steerable Catheter

The mechanical efficacy and MR-visibility of the steerable catheter were investigated in various simulated clinical scenarios. To assess MR-visibility, four catheters fabricated in-house were placed in a water-filled container. High-resolution MR images were acquired using a clinical 3-Tesla MR system with a 3D FLASH sequence. The resulting images (Fig. 5, A to C) clearly displayed the catheter shaft and tip markers, producing artifacts comparable in size to previous studies with guidewires *(38)*. The shaft artifacts had a mean diameter of $2.0 \pm 0.2$ mm, while the tip marker measured approximately 12.5 mm by 12.5 mm. Importantly, the inclusion of four MR-visible pull-wires for steering did not have a significant impact on the size of the artifacts. On average, the shaft artifacts had a thickness of 3.55 mm, ensuring excellent visibility in MR images.

To assess the catheter's four-way steering capability, we employed two projected planes that provided simultaneous visualization of a transverse plane at the distal-most point of the catheter tip and a frontal plane spanning the entire catheter length. This imaging approach allowed us to observe and document the catheter's precise movements in different directions, including left, right, up, and down. The results (Fig. 5D and movie S1) vividly showcased the catheter's exceptional maneuverability and its ability to navigate challenging anatomical pathways. Furthermore, through post-processing, a comprehensive illustration that clearly demonstrated the catheter tip's steering capability in all four directions provides a valuable tool for real-time guidance during procedures.

To validate the clinical applicability of our steerable catheter, phantom experiments were conducted using a soft silicone vascular phantom resembling a normal adult abdominal aorta. The phantom was filled with water, and under the guidance of real-time MRI, a novice operator successfully maneuvered the catheter into three primary target vascular structures: the right renal artery (RRA), left renal artery (LRA), and coeliac trunk. As showcased in Fig. 6 and movie S2, the probing procedure was further confirmed using contrast agent injection for enhanced visualization. These results highlight the feasibility and efficacy of our steerable catheter in navigating complex anatomical structures, even for operators with limited experience. Preliminary results on similar experiments can be found in *(28)*.

### *In vivo* Animal Study for Steerable Catheter and Active Tracking Catheter

To validate the clinical feasibility and performance of our steerable catheter, we conducted *in vivo* testing using a porcine model under general anesthesia. Real-time MRI was employed to monitor the catheter's position during the intervention, ensuring accurate guidance throughout the procedure. The cardiologist relied on a non-magnetic and RF-shielded in-room screen for real-time visualization of the catheter. Utilizing a pre-implanted femoral access sheath (10F), the catheter was successfully introduced into the arterial system. MRI displayed both the anatomy and the catheter, offering comprehensive coverage of the vascular system. Remarkably, the maneuverability of our catheter was effectively demonstrated by successfully bending the catheter tip in ventral direction to enable crossing of the aortic arch towards the heart (Fig. 7 and movie S3). Throughout the study, we adhered to the guidelines of the local ethics committee and regional council, ensuring the utmost care and compliance with animal welfare standards. These promising results underscore the potential of our steerable catheter to revolutionize medical interventions and contribute to improved patient outcomes.

Similar experiments were performed in a porcine model using an active tracking catheter. As shown in Fig. 8 and movie S4, the catheter is clearly visible under real-time magnetic resonance imaging. Both the location and shape of the catheter tip are depicted by the bright signal from the loop coil. In addition, the shaft of the catheter is visible as hyperintense signal that is picked up by the coaxial cable. Using the active tracking catheter, the left coronary artery (LCA) was successfully intubated. This allowed for a selective perfusion measurement of the myocardial segments supplied by the LCA. Therefore, gadolinium-based contrast agent was injected via the catheter under real-time imaging in a mid-ventricular short-axis view. The injection resulted in a selective signal enhancement in the septal, anterior and lateral segments of the left ventricle demonstrating the successful intubation of the LCA. It is worth noting that the active tracking catheter though not MR-safe by definition has never caused any adverse effects during more than 30 procedures.

## DISCUSSION

The versatility and precision of thermal drawing technology have garnered widespread recognition across various industries for their ability to produce fibers with precise cross-sectional geometries. While devices with high aspect ratios are commonly used in surgical procedures, the application of thermal drawing technology in the medical field remains largely unexplored. In this study, we effectively showcased the feasibility of employing thermal drawing technology in the production of catheter devices for medical applications, focusing specifically on the development of catheters for MR-guided endovascular intervention. Compared to the commonly used extrusion method, which can be complex and costly, thermal drawing presents a cost-effective and expeditious alternative.

The applicability of the technology spans across a diverse array of materials. Depending on the specific material, preform fabrication can be achieved through various means. For instance, this study demonstrates the utilization of direct machining of purchased material bars. Alternatively, more intricate preform fabricated components with complex geometries can be produced through techniques like 3D printing *(44)* and molding. This technique also allows the integration of multiple materials with different thermal and mechanical properties into a single preform, which is then thermally co-drawn to achieve the desired dimensions *(50, 51)*. The thermal drawing process also allows for precise adjustment of the cross-sectional area by manipulating the drawing speed, enabling the production of components with tailored dimensions as per the specific design requirements. Furthermore, this process has the capability to introduce rotational freedom, thereby enabling the creation of lumina with a helical structure. However, it is important to emphasize that careful temperature regulation during the heating phase and precise control of drawing velocity are crucial to ensure optimal heating and achieve the desired fiber diameter.

The feasibility of our proposed technology is demonstrated through the design and manufacturing of catheters for MR-guided intervention. MRI offers radiation-free imaging and provides morphological and functional data, thus showing promise for image-guided vascular navigation. The utilization of MR-guided intervention has gained significant attention, leading to investigations into the development of compatible essential instruments *(26)*. However, the selection of materials for these instruments remains constrained by considerations of mechanical properties and MR-visibility. In order to address this challenge, we have devised two distinct types of catheters for use in MR-guided endovascular intervention: a steerable catheter system featuring passive tracking, and a Tiger-shaped catheter equipped with active tracker.

The in-house fabricated catheter shafts exhibit similar mechanical properties to their commercial counterparts, reaffirming the efficacy of our manufacturing approach. However, the experimental methodology can be improved by conducting experiments in a controlled temperature and humidity environment. Additional research in a simulated intravascular environment, taking into account body temperature and moisture absorption, would provide valuable insights.

The steerable MR-safe catheter assembly underwent a comprehensive assessment to evaluate its functionality, and it successfully met the design requirements. Notably, the assembly demonstrated flexible four-way steering of the catheter tip, a reliable locking mechanism, effective fluid injection capability, and successful guidewire insertion. Both in-vitro and in-vivo experiments confirmed the compensatory impact of the helical lumen, effectively counteracting undesired motion under tendon driven. However, the implementation of the helical luminal structure also led to an increased tendon travel distance, resulting in heightened friction. While various efforts were made to eliminate backlash, achieving complete avoidance remains challenging. The knobs of the handle were specifically engineered with contours resembling those of gears, enabling more efficient automation. To ensure the visibility along the whole length of the catheter, passively MR-visible MaRVis rods *(52)* were implied as pull-wires to steer the catheter. These rods have been demonstrated to deliver sufficient pulling force for steering the catheter tip; however, kinks can happen when pushing forces were applied on them. Passive markers are inherently MR-safe and can provide visible artefacts indicating the two ends steerable section without significantly compromising the visibility of surrounding anatomical structures.

Active markers, in contrast, provide a strong positive signal at the marker locations. The combination of the catheters with an active tracking RF coil was very advantageous as it allowed to visualize the catheter tip with a positive contrast during the intervention. Positive contrasts are needed to identify the tip position and the catheter orientation; both parameters are vital to successfully advance a catheter into a branching vessel such as the coronary artery *(46)*. However, active tracking uses coaxial cables to guide the MRI signal from the tracking coil to the distal receiver which always entails the risk of device heating, especially when the device length in the MRI advances the resonance length in tissue. To overcome this risk several methods have been proposed such as integrated baluns *(53)* or transformers *(54)* – these additional structures might be more difficult to integrate during the extrusion process. Alternatively, the termination impedance of the circuit can be monitored leading to a reduced RF heating *(55)*, or the devices are operated at lower field strength MRI systems where heating is significantly less pronounced *(56)*.

Overall, the successful application of thermal drawing technology in the fabrication of catheter devices highlights its potential as a valuable manufacturing method in the medical field. By offering versatility in material integration and design flexibility, thermal drawing technology could revolutionize the way high aspect ratio medical devices are manufactured, offering effectiveness and customization that align with the dynamic needs of medical interventions.

## MATERIALS AND METHODS

### Multi-lumen tubing

Both PC and PEI polymers were selected as the material for the steerable catheters, using medical/life science grade (LSG) polymer rods (Mitsubishi Chemical Advanced Materials UK Ltd., Lancashire, United Kingdom) for the manufacture of the preform. Precise placement of the through-holes in the preform was ensured through peck drilling with driller of 6 mm. The thermal drawing process began by feeding the preform into the draw tower's three-zone tube furnace (TV-05 Solid Tubular Furnace, The Mellen Company, USA) at a downfeed speed of $v_{DF} = 1$ mm/min.

The preheating temperature in the top zone was set to 140 °C for PC and 220 °C for PEI; the middle zone temperature was adjusted above the glass transition temperature of PC ($T_g$ = 145 °C, set temperature = 240 °C) and PEI ($T_g$ = 215 °C, set temperature = 340 °C); bottom zone temperatures were set at 85°C for PC and 95°C for PEI. The fiber was monitored using a laser micrometer (Laserlinc, USA) to measure the outer diameter and a three-wheel tension sensor to measure its tension, which is ideally maintained at around 100g. The draw speed ($v_D$) is determined by applying the conservation of mass equation.

$$A_P * v_{DF} = A_F * v_D$$

Where, $A_P$ and $A_F$ are the cross-sectional area of the preform and the fiber.

The selected material for the active tracking catheter shaft was PC. The cross-section of the preform, depicted in fig. S1B, was prepared by drilling two holes (28.5 mm and 3 mm) into a PC rod. During the drawing process, temperatures were set to 140 °C, 240 °C, and 85 °C across the three zones, with $v_{DF}$ = 1 mm/min and $v_D$ = 225 mm/min, ensuring a consistent fiber diameter of 2.5 mm. A customized wire feed mechanism, featuring a wire spool and pulley, was used to guide and passively feed the wire of diameter 0.2 mm through the preform during drawing. As the preform was drawn, reducing the channel diameter, the wire was integrated into the fiber, allowing for the incorporation of metal wires using our existing fiber drawing platform.

The tip of the active tracking catheter is a single-lumen tubing made from COCe (E-14, TOPAS Advanced Polymer GmbH, Germany) through the thermal drawing. The preform (inner diameter = 32 mm, outer diameter = 40 mm) was created through vacuum compression molding *(29)*, at a heating temperature of 190 °C, pressure of 0.2 MPa, and time duration of 20 hours. After drawing (middle zone temperature = 210 °C, $v_{DF}$ = 1 mm/min, $v_D$ = 256 mm/min), the resultant single-lumen tubing has an inner diameter of 2 mm and outer lumen of 2.5 mm.

**FEA of helical lumen catheter**

FEA was carried out to investigate the uncontrolled tip steering caused by the 100 mm shaft being passively deflected by 5 mm (ANSYS 2020 R2). The tip and shaft material were set to PC with the following physical/mechanical properties: density = 1200 kg/m³, Young's modulus = 2380 MPa, Poisson's ratio = 0.399, tensile yield strength = 6.21 MPa. On the other hand, the pull wire was set to stainless steel with the following physical/mechanical properties: density = 7500 kg/m³, Young's modulus = 1.93e5 MPa, Poisson's ratio = 0.31 and tensile yield strength = 207 MPa. The pulling wire was fixed at both ends of the catheter and the intersection surface between the proximal end of the tip and the distal end of the shaft is also fixed. A passive offset of 5 mm was applied midway (50 mm) from the proximal end of the shaft.

**Post-processing of steerable catheter**

After drawing the fibers to the desired length, they underwent post-processing steps (fig. S2) to enhance their mechanical properties. Over-braiding was employed along its length, excluding the distal end, with Kevlar® 49 yarns (DuPont, Delaware, United States) creating a 1/1 diamond pattern *(57)*. This process is carried out using a vertical Maypole type braider (Herzog RU 2/16-80, HERZOG GmbH, Germany).

The hinge design was optimized based on the review article by Jelínek et al. *(58)*, and computational analysis using detasFLEX *(59)*. Micromachining of the distal end employed the LML-femto2000 ultrafast laser micromachining tool (Laser Micromachining, St. Asaph, United Kingdom), with a 343 nm ultraviolet femtosecond laser (Pharos SP-HP, Light Conversion, Vilnius,

Lithuania). The multi-lumen tube is fixed within a collet on a lathe stage (LaserTurn1, Aerotech, Inc., Pennsylvania, United States) during the notching procedure.

To affix a short segment of platinum-cured silicone tubing (SFM3-2050, Polymer Systems Technology Ltd, High Wycombe, United Kingdom) to the distal end of the catheter, two polyethylene heat shrink tubing pieces with a wall thickness of 6 μm (103-0042, Nordson MEDICAL, New Hampshire, United States) were utilized. This encapsulated assembly was further enhanced by applying an additional layer of hydrophilic polymer coating (Patented Hydrophilic Coating 8-3C, Coatings2Go, Massachusetts, United States). Lastly, two passive negative markers infused with iron microparticles are incorporated into the catheter assembly.

### Post-process of the active tracking catheter

A Tiger-shaped tip is achieved using a 3D-printed mold with a specific slot for the catheter tip, which is left in place for days to fully relax internal stresses, ensuring the tip retains its Tiger shape after removal. The resonance coil was then attached to the tip of the catheter and connected to the coaxial wire embedded in the catheter shaft. The COCe tip and the PC shaft were joined with a PTFE shrinking tube.

### Integration of steerable catheter

Details of the steerable catheter assembly are depicted in fig. S3. For visibility under MR images, MaRVis rods *(52)* were utilized as pull-wires within the catheter. Most of the catheter components were printed from a biocompatible Acrylonitrile Butadiene Styrene (ABS) using a Fused Deposition Modelling (FDM) printer (FORTUS 400mc, Stratasys Ltd., Eden Prairie, MN, USA). Plastic screws (PEEK and Glass Reinforced Resin Screws, Misumi Europa GmbH, Frankfurt, Germany), bearings (Linear Plain bearing GSM-0608-04, Igus, Northampton, United Kingdom), and the pinion (0.5 Mod*13 Tooth Metric Spur Gear In. Hostaform, Bearingboys, Norwich, United Kingdom) were excluded from the printing process.

The integration of the catheter involved inserting the catheter into the handle through a pin vice, passing a series of guide channels to suppress its lateral motion, and a pull-wire routing block that prevents crosstalk of the pull-wires and friction-caused damage. The shaft was secured in place by tightening the vice nut. Four polytetrafluoroethylene (PTFE) tubing (0.4*0.6, Iwase, Kanagawa, Japan) were threaded through the pull-wire routing block, extending through skived side holes that provide access to the four lateral channels of the shaft. Pre-lubricated pull-wires were fed through the PTFE tubing guide and advanced through the distal end of the catheter. Segments of fused silica capillary (TSP450670, CM Scientific, Keighley, United Kingdom) acting as stoppers were affixed using cyanoacrylate glue (Loctite 416, Henkel, Dusseldorf, Germany) at the tendon's distal end. The proximal ends of the pull-wires were equally tensioned and secured to their corresponding racks using a combination of a locking screw and Loctite 416. The racks were then held in place by the knob/pinion assembly, which was mounted to the handle using an external cover embedded with a mechanical lock. To minimize backlash, the pull-wire routing block was fixed using hot melt adhesive, effectively preventing unintended displacement of the pull-wires. Finally, a Luer lock port was connected at the proximal-most point of the catheter shaft.

### Benchtop Experimental setups

In order to assess the mechanical characteristics of the catheter shaft, including flexibility (fig. S6), pushability (fig. S7), torqueability (fig. S8 and fig. S9), steerability (fig. S10 and fig. S11), and the suppressive effect of the helical structure on undesired motion (fig. S12), a series of experiments were conducted using our custom-designed experimental setups. Materials and methods are available as supplementary materials.

### *In vitro* **and** *in vivo* **study**

All MRI experiments were performed on a clinical 3-Tesla system (Magnetom Prisma, Siemens). The visibility of the steerable catheter was first tested using a 3D FLASH sequence with the following parameters: echo time (TE), 2.2 ms; repetition time (TR), 5.0 ms; flip angle ($\alpha$) = 10°; field of view (FoV), 300*300*50 mm$^3$; matrix size, 240*240*40; bandwidth (BW), 400 Hz/pixel. MR images were acquired for four catheter samples immersed in water under these settings. The bending of one steerable catheter tip was imaged in real-time imaging using a 2D bSSFP sequence acquired in two orthogonal planes under settings of TE/TR = 1.3/3.0 ms, $\alpha$ = 43°, BW = 930 Hz/pixel, FoV = 300*56 mm², matrix size = 192*138, slice thickness (SL) = 15 mm.

A silicon phantom in the shape of a normal adult abdominal aorta (Elastrat Sarl, Geneva, Switzerland) was used for the accessibility investigation. Probing of the right and left renal arteries, as well as the coeliac trunk was imaged in real-time using a 2D FLASH sequence (TE/TR = 2.0/4.9 ms, TA/image = 251 ms, $\alpha$ = 11°, BW = 500 Hz/pixel, FoV = 300*256 mm², matrix size = 192*138, SL = 10 mm). To verify the successful intubation of the individual arteries, a bolus of diluted gadolinium contrast agent was injected through the catheter.

The steerable catheter and the Tiger-shaped active tracking catheter were tested in-vivo in two juvenile domestic landrace pigs. The experimental protocols received approval from the local ethics committee of Freiburg University and the regional council of Freiburg, Baden-Wuerttemberg, Germany (license numbers 35–9185.81/G-15/156 and 35-9185.81/G-16/78) and the experiments were conducted in accordance with FELASA and GV-SOLAS standards for animal welfare. The animals were anesthetized and mechanically ventilated throughout the experiments. A 10F arterial access sheath was introduced into the right femoral artery via ultrasound guidance prior to the MR exam. The animals were positioned in head first supine position with the heart at the iso-center and the 32-channel spine coil as well as an anterior 18-channel coil array were used for signal reception. Navigation of the catheters was guided using a real-time FLASH sequence (TE/TR = 1.3/3.4 ms, $\alpha$ = 10°, FoV = 289 mm², matrix size = 192*144, BW = 790 Hz/pixel, SL = 8 mm). First, the steerable catheter was tested in one animal. Therefore, the catheter advanced towards the aortic arch in straight configuration. Then, the tip was intentionally bent towards the ventral direction to enable to further advance the catheter across the ascending aorta to aortic root. The active tracking catheter was tested in the second animal. The catheter was advanced over a 0.018" guidewire (Terumo) until the catheter tip was positioned in the ascending aorta. Then, the guidewire was removed, and the catheter tip was maneuvered into the LCA.

The successful intubation was verified by performing a selective perfusion measurement. Therefore, a bolus of 1% gadolinium solution was injected via the catheter during imaging of the short-axis of the heart with an inversion recovery FLASH sequence (TE/TR = 1.0/2.0 ms, TI = 95 ms, $\alpha$ = 10°, FoV = 360*70 mm², matrix size = 192*106, BW = 1185 Hz/pixel, SL = 8 mm, single breath-hold).

**Acknowledgments:** MEMKA thanks David J. Williams and Paul Brown for their support in the manufacturing of the preforms. MEMKA also thanks Ning Liu and Anzhu Gao for their insightful discussions regarding the monolithic design of the steerable catheter. AM received support from the Berta-Ottenstein-Program for Advanced Clinician Scientists, Faculty of Medicine, University of Freiburg. TH, SR, AM, MB and CvzM members of SFB1425, funded by the DFG Project #422681845.

**Funding:**

Engineering and Physical Science Research Council UK grant EP/P012779 (GZY, EY)

Deutsche Forschungsgemeinschaft (DFG; German Research Foundation) research grant Project #492563001 (MB, AM) and #494790146 (SR)

**Author contributions:**

Conceptualization: MEMKA, LT, BT

Methodology: MEMKA, LT, TL, SR, TH, AM, KD, CvzM, BT

Investigation: MEMKA, LT

Visualization: MEMKA, LT

Funding acquisition: GZY, EY, BT

Project administration: BT

Supervision: BT

Writing – original draft: MEMKA, LT

Writing – review & editing: SR, MB, BT

**Competing interests:** The patent WO2022136869A1 is filed by the Imperial College Sci Tech & Medicine, published on 23 January 2020, with authors MEMKA, BT, and GZY on the patent. The authors declare that they have no other competing interests.

**Data and materials availability:** All data are available in the main text or the supplementary materials.

**Supplementary Materials**

Materials and Methods

Figs. S1 to S28

Tables S1 to S4

Movies S1 to S4

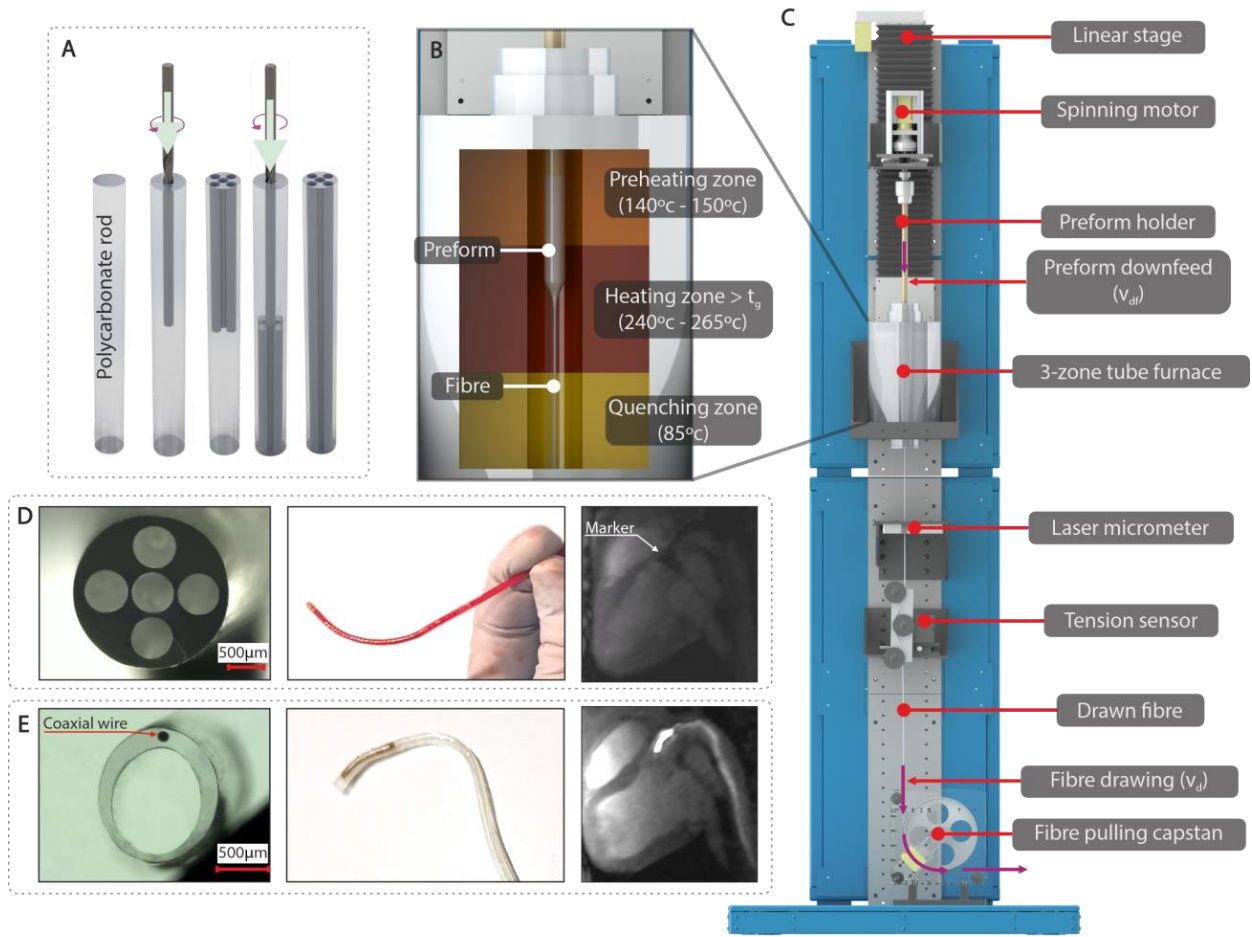

**Fig. 1. Fiber drawing tower and resultant catheters.** (**A**) CAD representation of the preform fabrication process for the steerable catheter: polymeric rod peck drilling performed from both ends. (**B**) Cross-sectional inset of the three-zone tube furnace, indicating the corresponding preheating, heating, and quenching temperatures. (**C**) CAD representation of the fiber draw tower. (**D**) Steerable catheter: (left) microscopic view of the cross-sectional structure; (middle) photograph of the catheter tips showcasing their unique designs; (right) in-vivo MR image of the catheters during testing in a living subject's heart. (**E**) Active tracking catheter: (left) microscopic view of the cross-sectional structure; (middle) photograph of the catheter tips showcasing their unique designs; (right) in-vivo MR image of the catheters during testing in a living subject's heart.

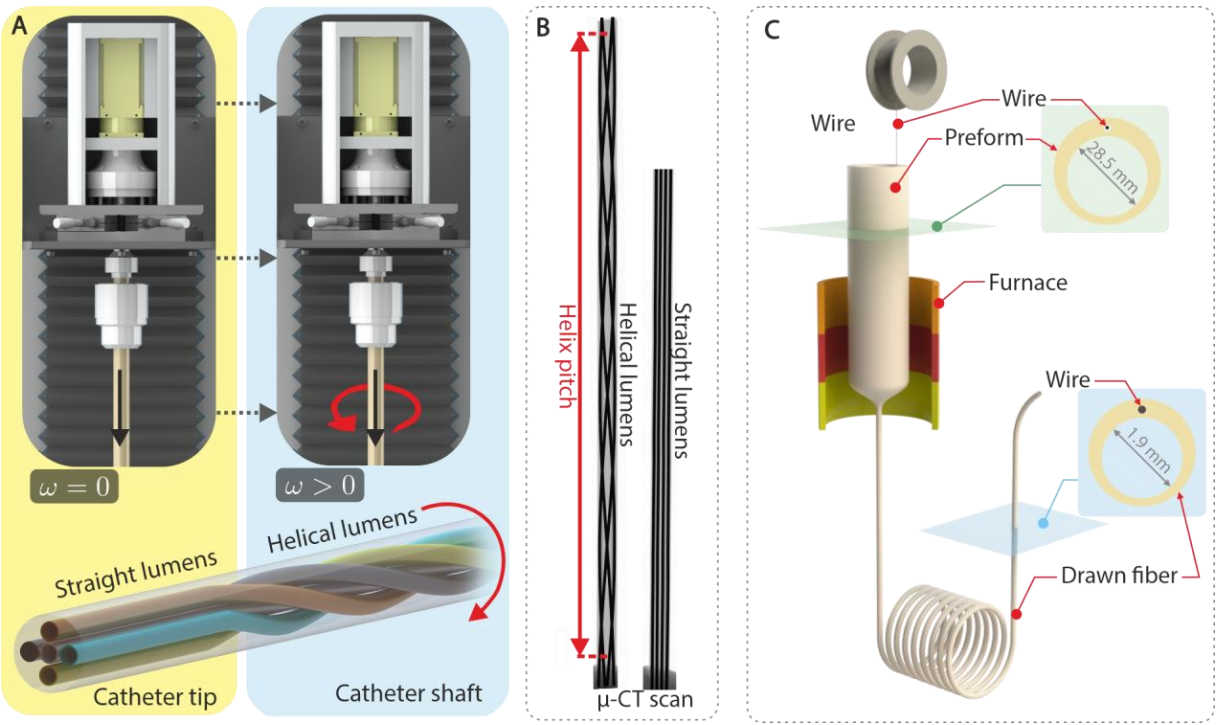

**Fig. 2. Preform twist and wire feeding in thermal drawing.** (**A**) Schematics illustrating the thermal drawing of multi-lumen tubing with helical lumina structure: Parallel lumina of the distal end fabricated by switching off the spinning motor (ω = 0), while helical lumina of the catheter shaft fabricated by switching on the spinning motor (ω > 0). (**B**) (left) μ-CT scan of multi-lumen tubing with parallel channels embedded with four 0.4mm nitinol wires for lumina visualization. (right) μ-CT scan of multi-lumen tubing with helical channels and a pitch size of 50mm. (**C**) CAD representation of the thermal drawing of multi-lumen tubing with wire feeding and optical image of the resultant fibers.

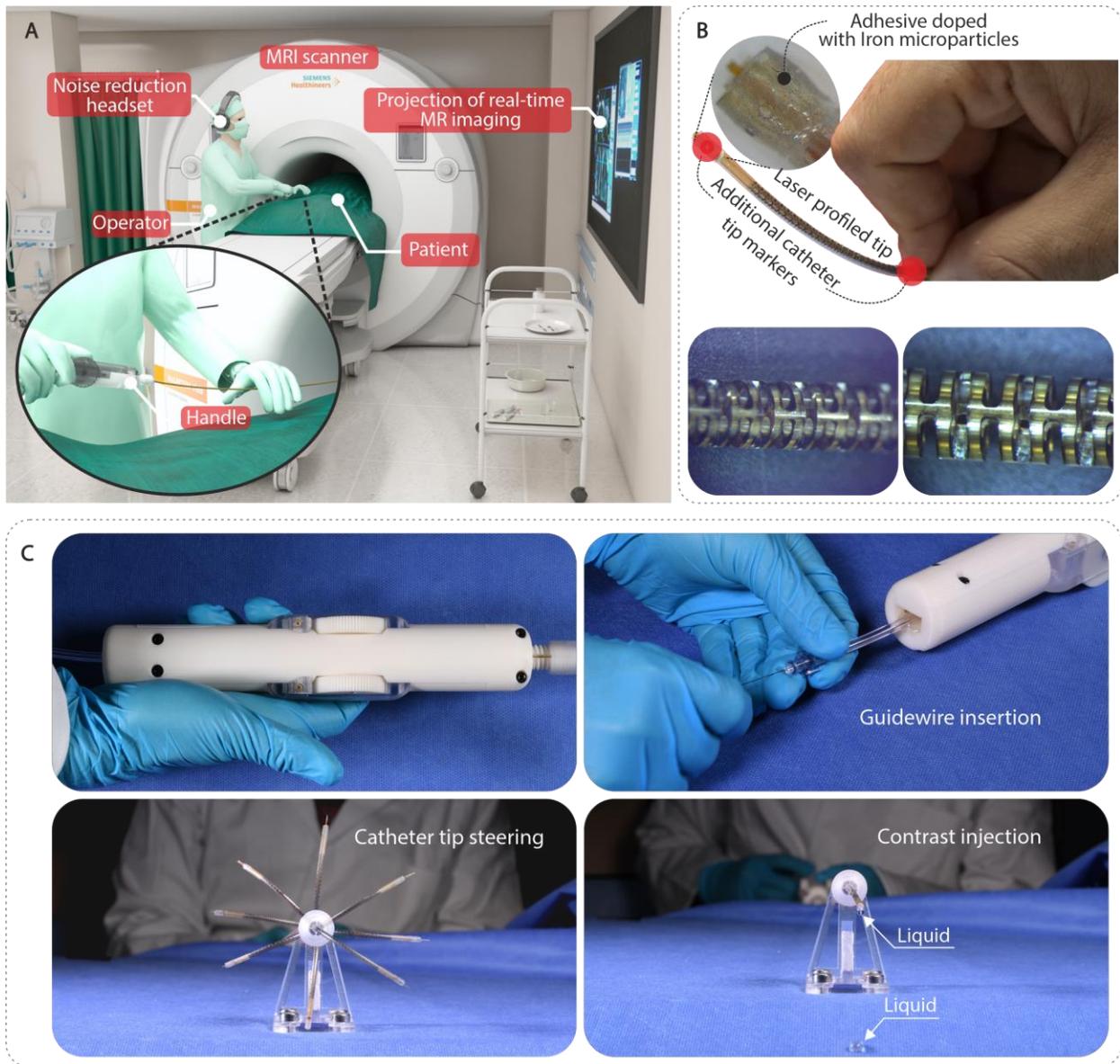

**Fig. 3. The tip details and functions of the steerable catheter system.** (**A**) CAD representation of the MR-guided endovascular intervention using the steerable catheter system in the MR suite. (**B**) (upper) Steerable catheter tip with labelled positions of additional tip markers. (lower) Detailed image of laser-profiled slots at the tip of the steerable catheter. (**C**) Top view of the catheter-handle assembly, and the basic functions of the assembled steerable catheter system: guidewire insertion from the back of the handle, catheter tip steering (superimposed images of 8 configurations) and contrast (fluid) injection.

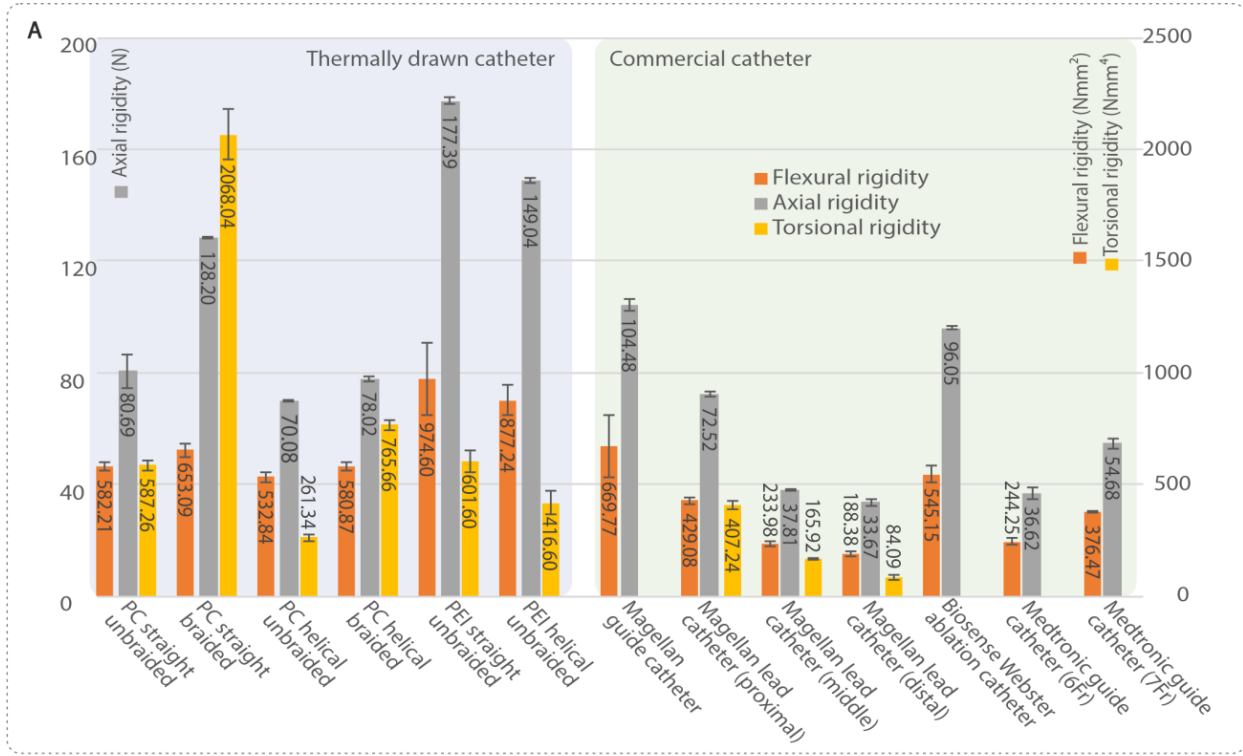

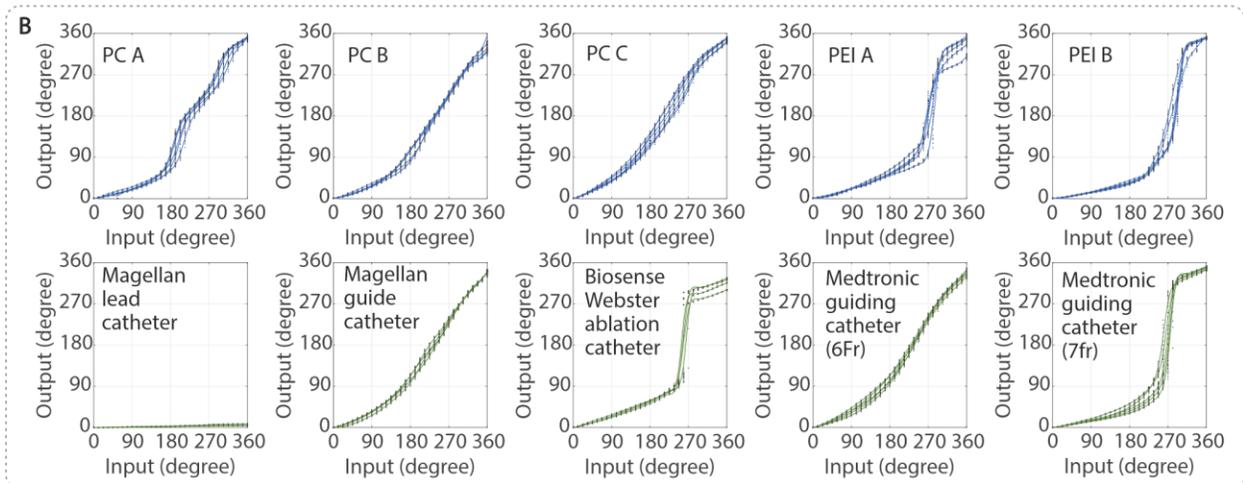

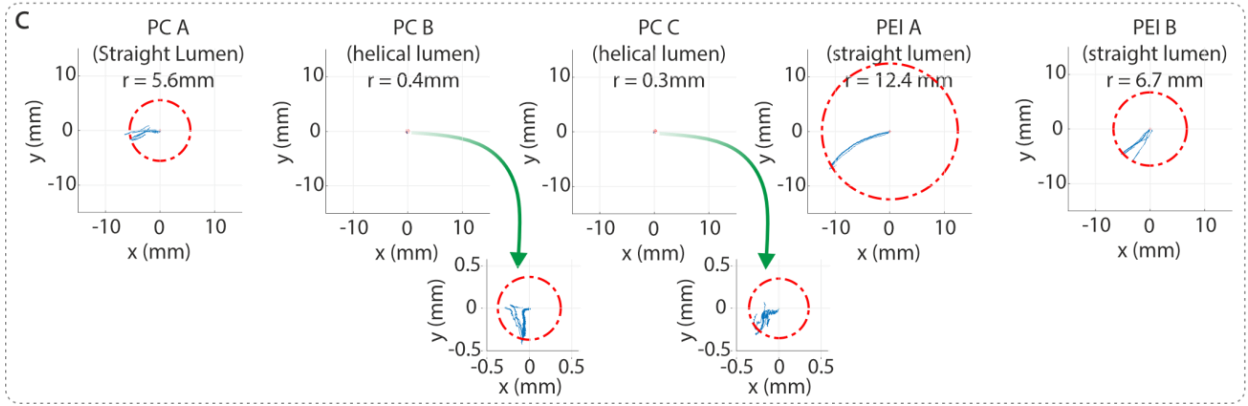

**Fig. 4. Mechanical characterization results for steerable catheter.** (**A**) Flexural rigidity (Experiment 1-1), axial rigidity (Experiment 2), and torsional rigidity (Experiment 3-1) of the catheter shafts. (**B**) Catheter shaft's angular response under rotary input (Experiment 3-2). (**C**) Uncontrolled deflection of the catheter tip resulting from a 90° catheter shaft deflection (Experiment 5).

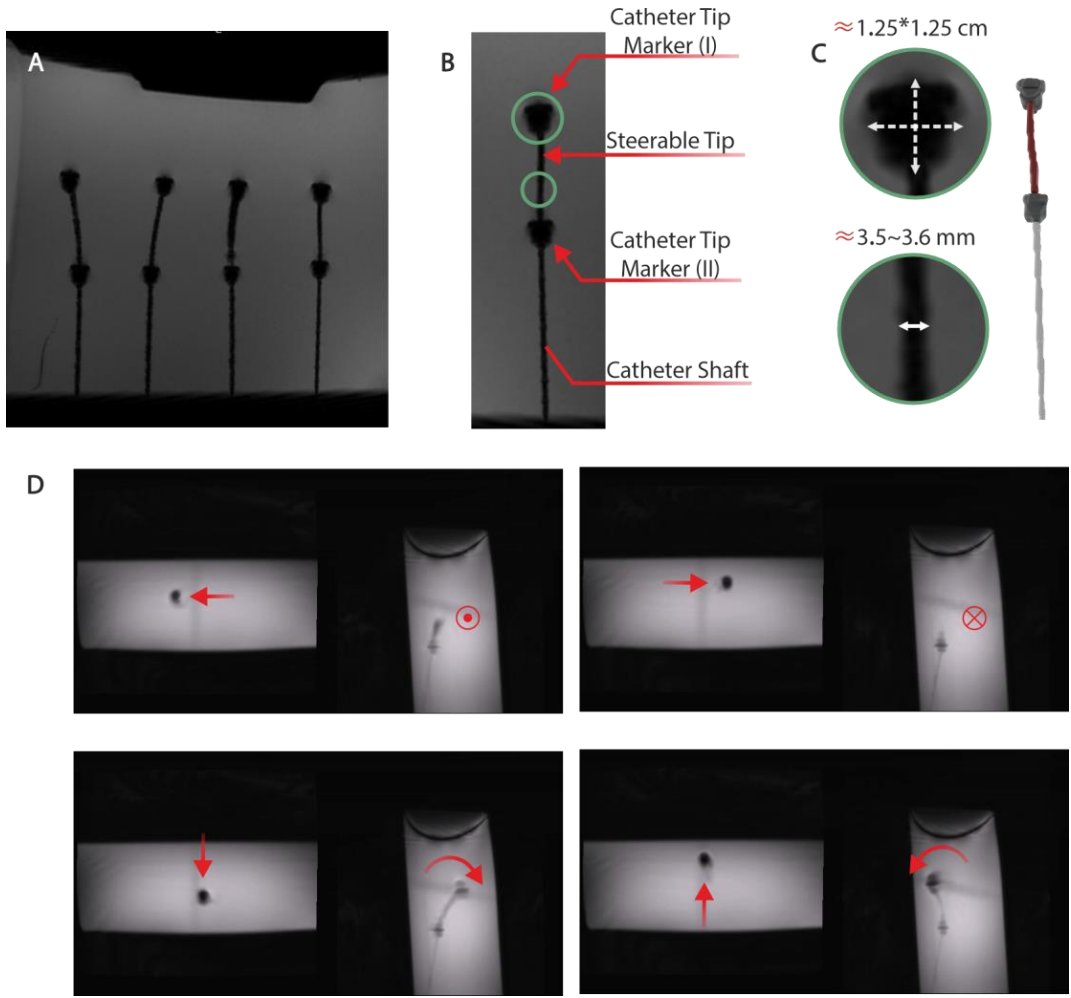

**Fig. 5.** *In vitro* **MRI test for steerable catheter.** (**A**) Imaging artefacts of four steerable catheters acquired using a 3D Flash sequence at 3 Tesla. (**B**) Labelling of the artefacts and inset of the catheter tip marker artefacts (upper) and catheter shaft artefacts (lower) with dimensions. (**C**) Reconstructed 3D model of the steerable catheter from the acquired scan. (**D**) Real-time imaging in the transverse and sagittal planes as the catheter is steered in all four directions.

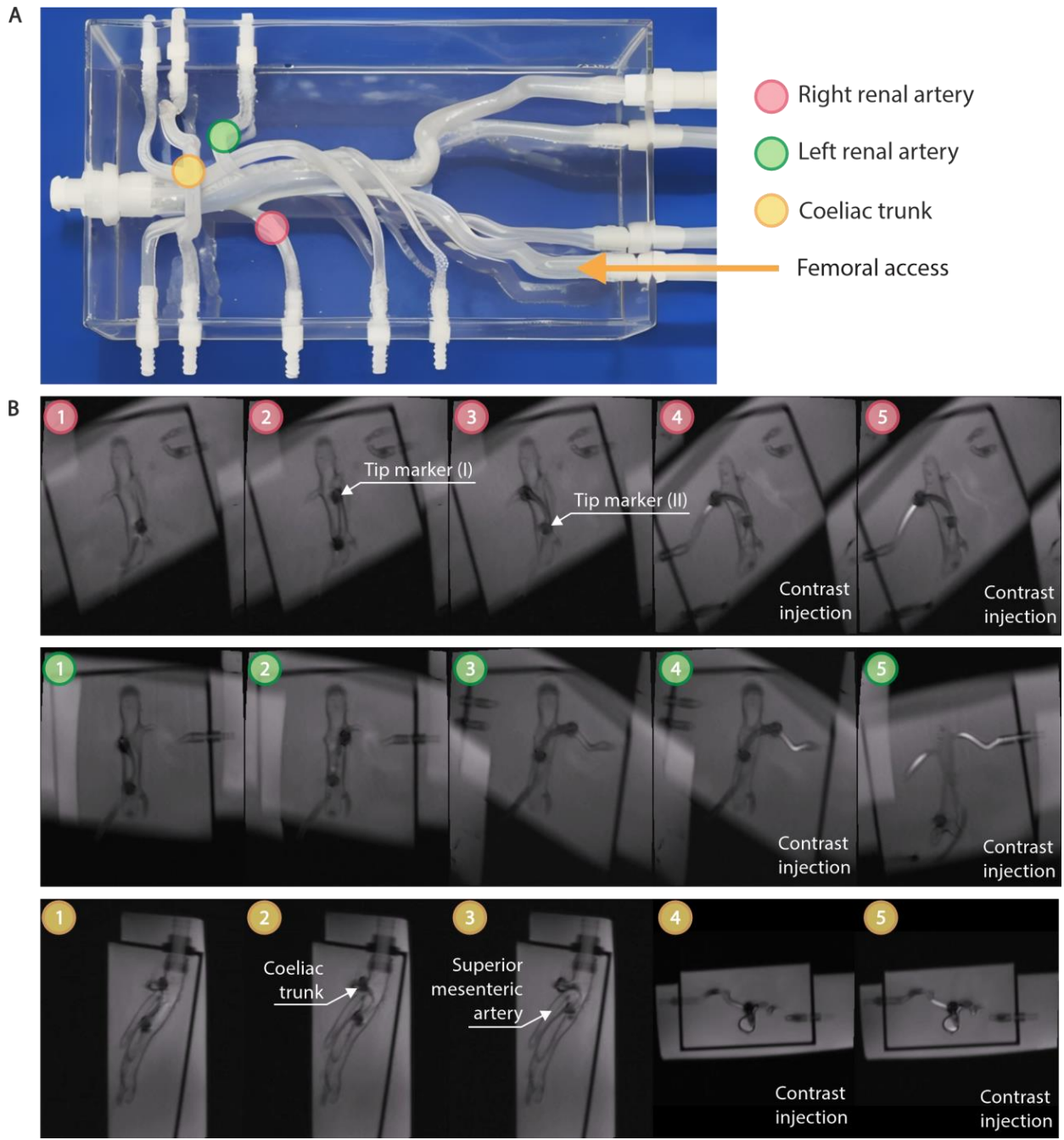

**Fig. 6. *In vitro* phantom study for steerable catheter.** (**A**) Abdominal phantom with left and right renal artery (LRA & RRA) and coeliac trunk artery target targets. (**B**) Image sequence illustrating probing of the LRA (green), RRA (red) and the coeliac (yellow) in an abdominal phantom using the steerable catheter under real-time MR guidance.

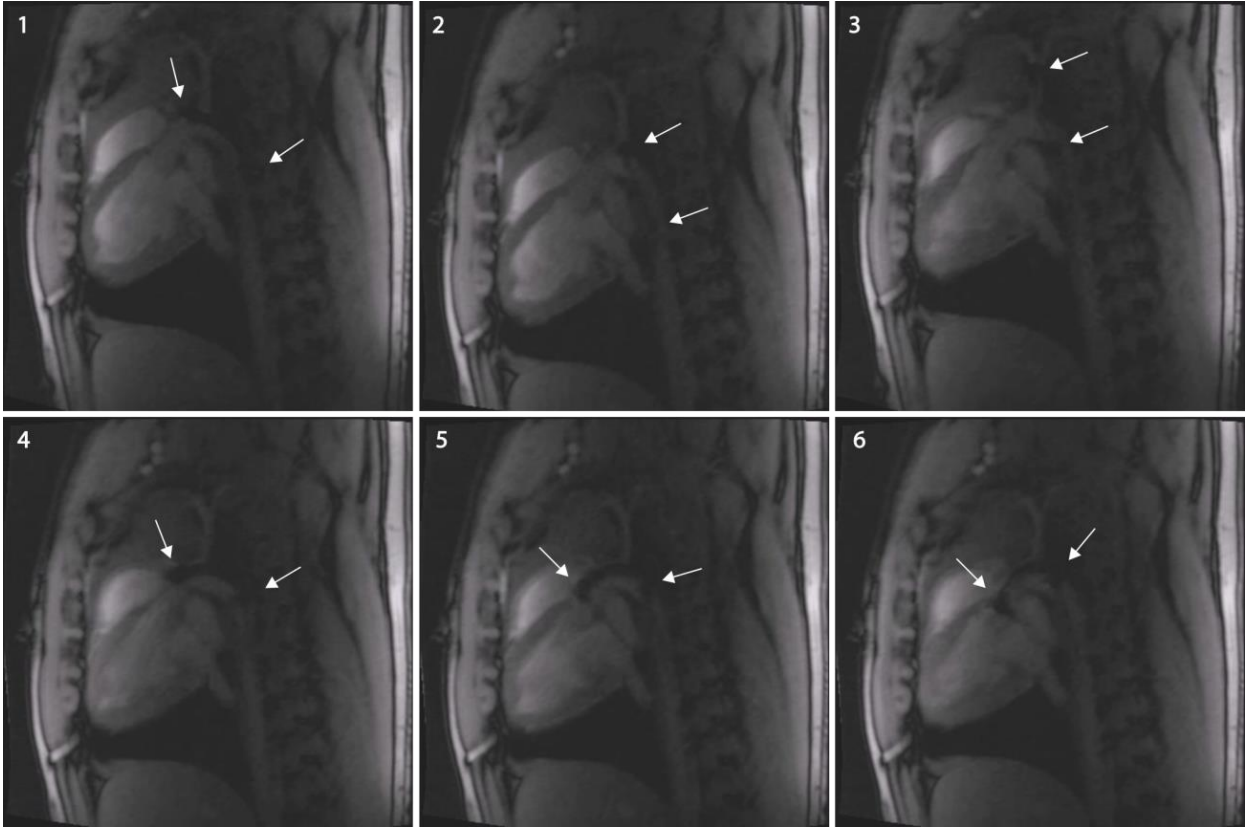

**Fig. 7. *In vivo* real-time MRI of steerable catheter manipulation in a clinical 3 Tesla system.** Arrows point at the two markers on the catheter tip. (1) Catheter advancing towards the aortic arch. (2-3) Catheter pointing towards the carotid artery. (4-5) Catheter steering towards the aortic arch. (6) Catheter crossing the aortic arch towards the aortic root.

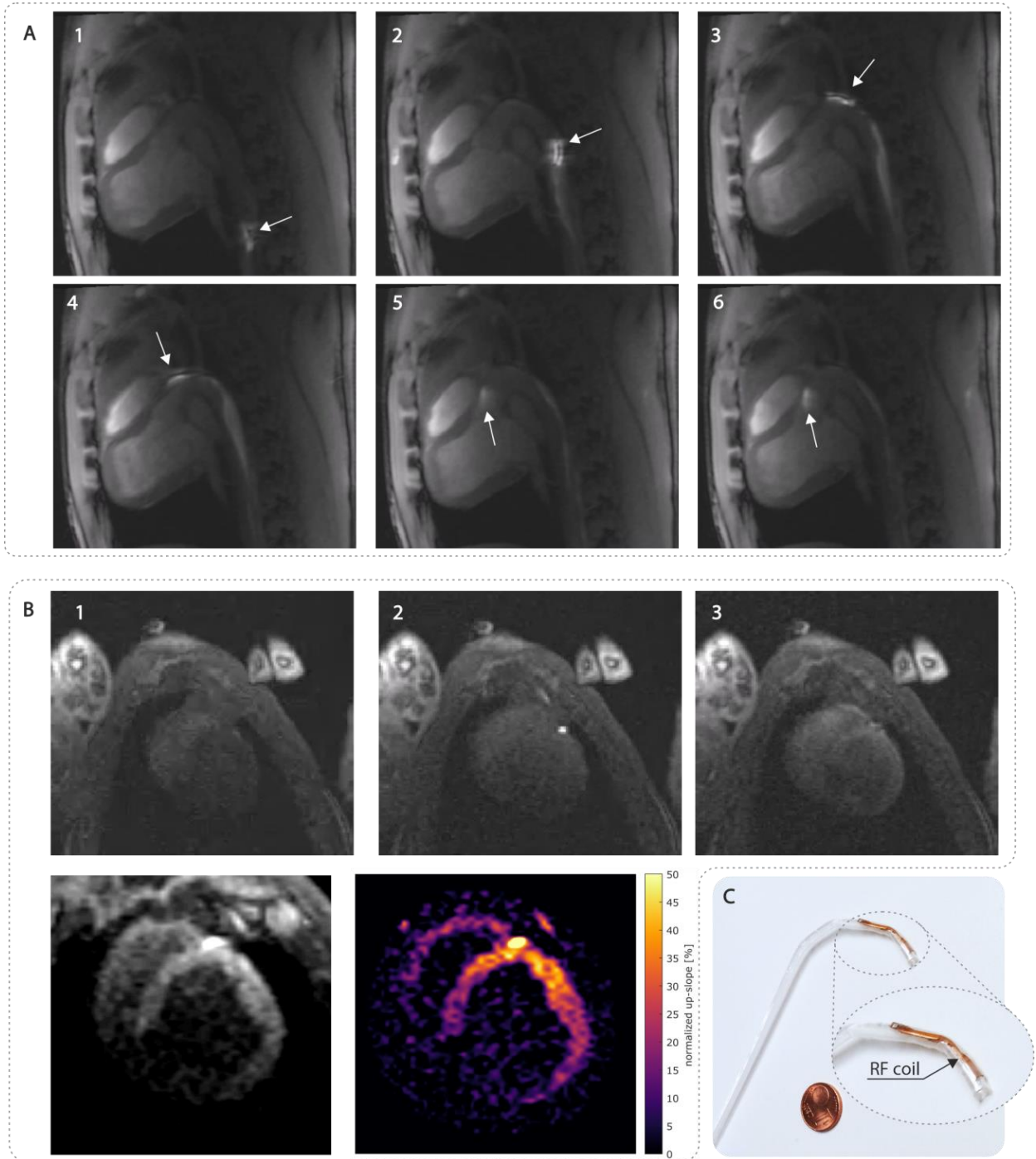

**Fig.8. *In vivo* real time MRI of active tracking catheter and photograph of tip details.** (**A**) Catheter crossing the aortic arch with arrows pointing at the catheter tip. (**B**) Intracoronary injection of gadolinium-containing contrast for selective perfusion, enabling semi-quantitative mapping by normalizing values to the up-slope in the left anterior descending artery. (**C**) Active tracking catheter tip with labelled positions of RF coil.